\documentstyle[11pt,epsfig]{article}
\columnwidth\textwidth

\newcommand{\avX}{\mbox{$\langle X\rangle$}}
\newcommand{\avXt}{\mbox{$\langle X(t)\rangle$}}

\begin{document}
\begin{flushright}
IFIC/FTUV-9831\\
hep-ph/9807361\\
\end{flushright}

\begin{center}
\Large{\bf Finite dimensional systems with random external fields
 and Neutrino propagation in fluctuating  media.}
\end{center}

\begin{center}
E. Torrente-Lujan. \\
IFIC-Dpto. Fisica Teorica. CSIC-Universitat de Valencia. \\
Dr. Moliner 50, E-46100, Burjassot, Valencia, Spain.\\
e-mail: e.torrente@cern.ch\\
\end{center}

\begin{abstract}
We   develop the general 
 formalism for the study of neutrino propagation in presence of
stochastic media. This formalism  allows 
 the systematic derivation of evolution equations for averaged 
quantities as survival probabilities and higher order distribution moments. 
The formalism applies equally to 
 any finite dimensional Schroedinger equation in presence of a 
stochastic external field. 
New integro-differential equations
 valid for finite 
correlated processes are obtained for the first time. 
For the
particular case of  
 exponentially correlated processes a second order ordinary equation is 
obtained.
As a consequence, the Redfield equation valid
 for Gaussian delta-correlated
noise is rederived in  a simple way. 
The formalism, together
with the quantum correlation theorem is applied  to the computation of  
higher moments and correlation functions of practical interest in 
forthcoming high precision neutrino experiments.
It is shown that equal and not equal time correlators follow similar differential equations.

\end{abstract}

\vskip 1cm

PACS numbers: 14.60.Pq; 13.10.+q; 13.15.+g;  14.60.Gh; 96.60.Kx; 02.50.Ey; 05.40.+j;  95.30.Cq;  98.80.Cq.  \\
{\bf Key words:} neutrino, neutrino mixing, neutrino interactions, stochastic processes.

\newpage

\section{Introduction.}

Neutrino oscillations in the presence of matter and magnetic fields have been
an area of intense interest for a long time. In many astrophysical
 situations the 
matter density and/or magnetic field 
may fluctuate  around a mean value. In some scenarios
it is expected that these fluctuations are very strong, for example 
when dealing with the 
magnetic fields that the neutrino encounters while propagating 
in the convective regions of the Sun. 

The case of neutrino spin precession in a noisy magnetic field 
with a $\delta$-correlated Gaussian distribution was 
considered initially by \cite{nic2} for neutrinos in vacuum.
In \cite{lor1}, a differential equation (a Redfield equation) 
for the averaged matrix density
was derived for the case in which the random noise 
was taken to 
be again a delta-correlated  Gaussian distribution.
Numerous works where approximate or numerical approaches are used
 to study  
the implications of matter random perturbations 
upon the SNP problem, Supernova dynamics or others 
are
avalaible (among them:
\cite{enqse,lor2,tor2,nun2,bur1,sem5,enq3} 
 see introductions in \cite{lor1,bur1}  for a more extensive review).

In this work 
we will  develop the general 
 formalism for the study of neutrino propagation in presence of
stochastic media. This formalism will  allow 
 the systematic derivation of evolution equations for averaged 
quantities as survival probabilities and higher order moments of them. 
New integro-differential equations
 valid for finite 
correlation processes are obtained for the first time. 
For exponentially correlated processes a second order ordinary equation is 
obtained as a consequence.
The Redfield equation valid
 for Gaussian delta-correlated
noise is rederived in  a simple way as a particular case: it is 
obtained as the zero-order term of a asymptotic expansion in 
 the inverse  of the correlation length.
 In the context of neutrino oscillations,
it is shown that the presence of matter noise induces the appearance  of 
a effective complex part in the matter density. The presence of magnetic 
field noise affects however in a qualitatively different, simpler, way. 
The range of validity of the $\delta$-correlated model is checked in 
a realistic   example corresponding neutrino oscillations in random 
 solar matter.
Finally the formalism, together
with the quantum correlation theorem will be applied 
to 
higher moments and correlation functions of practical interest specially in 
forthcoming high precision neutrino experiments.

The formalism may be applied to any 
quantum system governed by similar Schroedinger equations: equations where
a stochastic function appears multiplicatively in some part of the 
Hamiltonian. For simplicity of notation, in this work we will   always
deal 
 with Hamiltonians which are finite dimensional operators in some 
Hilbert space,  the equations will be obviously of the 
same type  for general systems
described by infinite dimensional 
Hamiltonians.

\section{The General Case.} 
Let's consider a general system  whose evolution is described by 
the linear random Schroedinger  equation
\begin{eqnarray}
i\partial_t X &=& \rho(t) L(t) X\; ; \quad X(0)=X_0
\label{e6101} 
\end{eqnarray}
where: $X$ is a vector (i.e. a multiflavour wave function) 
or a matrix (i.e. a density matrix) of arbitrary  dimension.
Eq.(\ref{e6101}) is defined in a interaction representation, 
 any additive non-stochastic term has been solved for  and absorbed in the definition of $X(t)$ and $L(t)$.
$L(t)$ is a 
general 
linear operator.
It can be an ordinary matrix  if $X$ represents a wave function. If $X$ is a density matrix, $L(t)$ is   a commutator:
\begin{equation}
L_A(t) X\equiv [A(t), X]
\end{equation}
Any other linear operator is admissible: obviously it is  
always possible to define
an enlarged vector space where the action of $L$ is represented by a matrix.  

We will suppose that 
 $\rho(t)$ in Eq.(\ref{e6101})  
is a scalar Gaussian process completely determined 
by its  first  two moments, which, without loss of generality can be taken as:
\begin{eqnarray}
\langle \rho(t)\rangle=0; & & \langle\rho(t)\rho(t')\rangle=f(t,t')
\end{eqnarray}
In other terms, 
$\rho(t)$  is 
characterized completely by the measure 
\begin{eqnarray}
[d\rho] \exp -\frac{1}{2}\int_{-\infty}^\infty \rho(t)f^{-1}(t,t')\rho(t')
 dtdt'. 
\label{e3bb}
\end{eqnarray}

An important particular case is when the process is $\delta$-correlated, 
 the correlation function is of the form: 
\begin{eqnarray}
f(t,t')=\Omega^2 \delta(t-t').
\label{e7001}
\end{eqnarray}
 A  convenient and practical way of parameterizing a 
 correlation function with a finite 
correlation length is to use an exponential function:
\begin{eqnarray}
f(t,t')=\Omega^2 \epsilon \exp\left (-\epsilon\mid t-t'\mid\right )
\label{e6105}
\end{eqnarray}

In this case the correlation length is defined as  
$\tau=1/\epsilon$. The expression (\ref{e7001}) is 
reobtained letting  
$\tau\to 0 $ or  $\epsilon\to \infty$ in Eq.(\ref{e6105}).

$X(t)$, the solution to Eq.(\ref{e6101}) is a stochastic function.
The objective of this work is to obtain in a systematic way equations for
its ensemble average and higher moments. 
Let us remark that 
for us Eq.(\ref{e6101}) is purely phenomenological, we suppose that is the
result of a more complete microscopic analysis which can account for  the
randomness of $\rho(t)$ (see for example \cite{bur1}).

To obtain a differential equation for the average of $X$ we make
use of the following well known property: For any Gaussian process $\rho(t)$ characterized by a $\delta$-correlation function   
(\ref{e6105}) and  any functional $F[\rho]$ 
we have the following relation (\cite{zinnj}):
\begin{equation}
\langle F[\rho] \rho(t)\rangle=\Omega^2 \left\langle\frac{\delta F[\rho]}{\delta \rho(t)}\right\rangle.
\end{equation}
For a Gaussian process with arbitrary correlation function (Eq.(\ref{e3bb})) 
we have the general relation:
\begin{equation}
\langle F[\rho] \rho(t)\rangle=\int d\tau \left\langle\rho(t)\rho(\tau)\right\rangle\left\langle\frac{\delta F[\rho]}{\delta \rho(\tau)}\right\rangle.
\label{e5201}
\end{equation}

 Let's consider now the evolution operator $U$  for a particular realization of
Eq.(\ref{e6101}); by definition:
\begin{equation}
X(t)=U(t,t_0) X_0.
\label{e5202}
\end{equation}
The operator $U(t,t_0)$ is a functional of $\rho$ and it has  the following formal expression in terms of a time ordered exponential:
\begin{equation}
U(t,t_0)=T \exp -i \int_{t0}^t\ \rho(\tau) L(\tau) d\tau .
\label{e9003}
\end{equation}
The functional derivative of $U$ with respect  to $\rho(t)$ can be computed by direct methods differentiating term by term the series expansion for Eq.(\ref{e9003}).
The result is:
\begin{equation}
\frac{\delta U(t,t_0)}{\delta \rho(\tau)}=-i L(\tau) U(\tau,t_0), 
\quad t_0<\tau<t.
\label{e5203}
\end{equation}

In order to obtain a differential equation for $\avX$ 
we observe that
\begin{equation}
i\langle\partial_t X\rangle=i \partial_t\avX=L(t) \langle\rho(t) X\rangle.
\end{equation}
Combining together Eq.(\ref{e5201},\ref{e5202},\ref{e5203})  we obtain
easily 
that  the evolution equation for the ensemble average is in the general case a 
integro-differential equation given by:
\begin{eqnarray}
i\partial_t \avXt&=&-i\int_0^t dt'\ f(t,t') L(t)L(t')\langle X(t')\rangle,\quad  \langle X(0)\rangle =X_0
\label{e6102}
\label{e9004}
\end{eqnarray}
This equation, which is exact and of very general validity, is the equation we were looking for and one of the main result of the present work.
Note that it has been obtained previously  integro-differential equations, valid for particular cases or in particular limits, using heuristic 
ad-hoc arguments (for example in \cite{enq3}). The derivation of  Eq.(\ref{e6102}) done here  
is the rigorous justification for such approaches.

\section{The particular $\delta$-correlated case.}
For the particular case where the correlation function is of exponential 
type, a second order ordinary differential equation can be obtained as we will see below. 
In the other hand,  
for the simpler $\delta$-correlated case
the evolution equation (\ref{e6102}) becomes the 
purely differential equation:
\begin{eqnarray}
i\partial_t \avXt&=&-i \Omega^2 L^2(t)\avXt.
\label{e6103}
\label{e9005}
\end{eqnarray}
Taking $L$ as a commutator,
this last equation coincides with the Redfield equation derived by 
\cite{lor1}. Note that the effective 
"Hamiltonian" appearing in the second part of Eq.(\ref{e6103}) is not anymore 
hermitic ( an example of a fluctuation-dissipation effect).

In practical cases of interest for the neutrino oscillation problem, the 
original equation for the density matrix  is of 
the slightly simpler form:
\begin{eqnarray}
i\partial_t X &=& \left [H_0(t)+\rho(t) g(t) H_1,  X\right ]\; ; \quad X(0)=X_0
\label{e6101c} 
\end{eqnarray}
where we have written explicitly the commutator. $\rho(t)$ is a stochastic function as before and $g(t)$ an arbitrary
scalar function. $H_0,H_1$ are Hamiltonian matrices, the former contains the 
average part of $\rho(t)$:
$H_0\equiv H_0'(t)+\rho_0(t) H_1$, with $\rho_0=\langle\rho\rangle$. 
The latter is supposed to be time independent.

For the problem described by Eq.(\ref{e6101c})
 the corresponding Redfield equation for the averaged 
density matrix  is of the form
\begin{eqnarray}
i\partial_t \avX
&=& [H_0(t), \avX]-i \Omega^2 g^2(t)[H_1,[H_1,\avX]].\\
&\equiv& H_0^- \avX-\avX H_0^++2 i \Omega^2 g^2(t) H_1 \avX H_1
\label{e6107c}
\end{eqnarray}
where in the last line the following  effective Hamiltonians were defined:
$$H_0^\pm=H_0\pm i g^2(t) H_1^2.$$

The solution of what is called the ''coherent'' part of Eq.(\ref{e6107c}) is accomplished 
by defining the average evolution operator:
\begin{eqnarray}
\langle U^\pm\rangle&=&T \exp \int d\tau H_0^\pm (\tau); \quad \left\langle U^-\right\rangle=\left\langle U^{+}\right\rangle^{\dagger}.
\end{eqnarray}
The   coherent part of the density matrix is then:
 \begin{eqnarray}
\avX_{coh}&=&\left\langle U^- \right\rangle X_0 \left\langle U^{+}\right\rangle
^{\dagger}.
\label{e4603}
\end{eqnarray}
Defining a new ''coherent'' interaction representation
by the relations: 
 \begin{eqnarray}
H_P&=&\langle U^-\rangle^{-1} H_1 \langle U^-\rangle, \quad 
H_Q=\langle U^-\rangle H_1 \langle U^-\rangle^{-1},
\quad
\avX_I=\langle U^-\rangle \avX_{coh} \langle U^-\rangle^{-1},
\label{e4602}
\end{eqnarray}
the resolution 
of the original equation is equivalent to the resolution 
 of the following one:
 \begin{eqnarray}
i\partial_t \avX_{I}&=&2 i g^2(t) \Omega^2 H_P \avX_I H_Q.
\label{e4601}
\end{eqnarray}

There are some  important particular practical 
cases where  Eq.(\ref{e9005}) can be
solved or simplified considerably taking into account the algebraic
 properties of a specific  $L$ (in what follows $k(t)$ is  always a 
scalar function):

{\bf A.} Let us suppose that $L$ is such that $L^2(t)=k(t) L(t)$. 
 This case appears in the computation of the average wave function
with matter density noise.
 Eq.(\ref{e9005}) reduces to: 
$$i\partial_t \avX=-i \Omega^2 k(t) L(t) \avX.$$
The averaged equation is similar  to the original one, the non-random part 
of the density is
``renormalized'' acquiring an imaginary term:
$$\rho\rightarrow \rho_0-i \Omega^2 k.$$
This is the density which will appear in the coherent effective Hamiltonians
$H_0^\pm$.
This case was discussed already in \cite{tor2}.  Enquist and Semikoz (\cite{enqse}) saw numerically that the net effect in the averaged survival probability 
of the same system is a precession reduction, similar to the effect of a larger constant matter density. It can be argued (Eqs. (\ref{e4603}) through (\ref{e4601})) that 
the complex renormalization defined before appears also, but this time approximately, if we compute average probabilities. Under this perspective the reduction in neutrino precession can be explained as a smoothening of the MSW resonance
induced by the imaginary term.  It is expected from here that $\delta$-correlated
 matter noise have importance 
only if applied over a MSW resonance region.

{\bf B.} The case where $L^2(t)=k(t) I$ with $I$   the identity matrix
 appears in the computation of the averaged neutrino 
wave function under noisy magnetic spin-flavour precession.  
The resulting equation
can be integrated trivially (to be compared with the previous case):
$$\langle X(t)\rangle=\exp -\Omega^2\int_0^t d\tau k(\tau)  \  X_0. $$
The average wave oscillation is damped by a factor equivalent to the one first calculated
by Nicolaidis (\cite{nic2}). This damping manifest itself also when computing
the average density matrix from Eq.(\ref{e4601}). 
From these difference of behavior with respect case (A) it is 
expected  that magnetic field noise can affect even if applied far from any 
resonance region.

{\bf C.} The case where  $L^4(t)=-k(t) L^2(t)$ appears in the 
computation of full
average density matrix with both, matter density or magnetic noise.
We can obtain in this case the ``conservation law'':
$$\left (1-\Omega^2 k(t)\right )L^2(t) \partial_t \avXt=0.$$
$L^2(t)$ is not invertible because the operator $L(t)$  has a 
zero eigenvalue. The previous expression
has proved to be of practical importance in  
some concrete numerical applications (\cite{tor3,tor4,tor5}).

\section{An Asymptotic Expansion for Exponentially correlated systems.}
 
We will see now  how the Eq.(\ref{e9005}) can be obtained 
as a limiting
particular case when the correlation length tends to zero.
For this purpose we use 
an exponential correlation function as Eq.(\ref{e6105}),  
the integro-differential 
evolution equation becomes in this case:
\begin{eqnarray}
i\partial_t \avXt &=&-i \Omega^2  \ \epsilon\exp(-\epsilon t)\int_0^t 
dt' \exp(\epsilon \tau) L(t)L(t') \langle X(t')\rangle.
\label{e6106}
\end{eqnarray}

Let us   compute the asymptotic expansion of the second term of 
Eq.(\ref{e6106}) valid for $\epsilon$ large, the following 
 expansion is valid  for any function $g(t)$ 
\begin{eqnarray}
h(\epsilon)&\equiv& \epsilon \exp(-\epsilon t)\int_0^t d\tau\ \exp(\epsilon 
\tau) g(\tau)\sim g(t)-\frac{g'(t)}{\epsilon}+\frac{g''(t)}{\epsilon^2}+\dots
\end{eqnarray}
Inserting this expression in Eq.(\ref{e6106}), we obtain the following 
expansion in 
powers of $\epsilon$:
\begin{eqnarray}
i\partial_t \avX&=&-i \Omega^2 L^2(t)\langle X\rangle+i \frac{\Omega^2}{\epsilon} L(t) 
\partial_t\left ( L(t)\avX\right )+o\left(\frac{1}{\epsilon^2}\right )
\end{eqnarray}
To leading order in $1/\epsilon$, we recover the expression corresponding to 
the $\delta$-correlated case. At next-to-leading order we get  
finite-correlation correction terms: 
\begin{eqnarray}
i\partial_t \avX&=&-i \Omega^2 L^2(t)\avX+i\frac{\Omega^2}{\epsilon}L(t)L'(t)
 \avX+i 
\frac{\Omega^2}{\epsilon}L^2(t)\partial_t \avX
\end{eqnarray}
or equivalently
\begin{eqnarray}
\left(1-\frac{\Omega^2}{\epsilon} L^2(t)\right )\partial_t \avX=\left (-\Omega^2 
L^2(t)+\frac{\Omega^2}{\epsilon}L(t)L'(t)\right ) \avX
\end{eqnarray}

We 
finally get the following differential equation valid to order $1/\epsilon$, 
 making the supposition that the operator which multiplies 
the first term is invertible, 
\begin{eqnarray}
\partial_t \avX&=&\left (-\Omega^2 
L^2(t)+\frac{\Omega^2}{\epsilon}L(t)L'(t)+\frac{\Omega^4}{\epsilon}L^4(t)\right 
) \avX
\label{e5205}
\end{eqnarray}

This equation can be used for finite, but relatively large, 
correlation lengths. We see that, up to this degree of approximation, not only 
the ratio level of noise to correlation length ($\Omega^2/\epsilon$) is 
important. We have different regimes 
according to the value of $\Omega^2$. 
For low noise amplitude ($\Omega^2<<1$) the first term will 
be more important. For strong noise ($\Omega^2>>1$) 
the second one, proportional 
in this case to $L^4$, will dominate.

We have pointed out previously that in the 
 practical cases of interest for neutrino oscillation problems the relation
$L^4(t)=k(t) L^2(t)$ holds, with $k(t)$  positive. If in addition  the term $L'(t)$ can be neglected, as indeed 
happens in some occasions, Eq.(\ref{e5205}) becomes:
\begin{eqnarray}
\partial_t \avX&=&-\Omega^2\left (1 -\frac{\Omega^2}{\epsilon} k(t) \right )L^2(t) 
 \avX
\label{e5205b}
\end{eqnarray}
In this case and  within this  level of approximation 
the  net effect of the presence of a finite correlation length is visible: 
it amounts to a  rescaling and consequent reduction of the noise parameter $\Omega^2$.

\section{An Exact Differential Equation for Exponentially Correlated 
Systems.}
 
In contrast with the 
approximate approach used in the previous section, we can 
derive in fact a simple, ordinary second order differential equation for the case of exponential 
correlation.  
Let's suppose that the original equation is of the same decomposable type as
the one appearing in Eq.(\ref{e6101c}) but let's include other cases using the 
general notation:
\begin{eqnarray}
i\partial_t X &=& \left (L_0(t)+\rho(t) g(t) L_1 \right )X\; ; \quad X(0)=X_0
\label{e6101b} 
\end{eqnarray}
with $L_0,L_1$ general linear operators as before, the latter 
time-independent.
In this case Eq.(\ref{e6102}) is of the form
\begin{eqnarray}
i\partial_t \avX= L_0(t) \avX-i \Omega^2 \epsilon L_1^2 e^{-\epsilon t} g(t)\int_0^t dt' g(t') e^{\epsilon t'} \langle X(t')\rangle.
\label{e6107}
\end{eqnarray}
Differentiating the equation once and after some simple algebra we obtain the following ordinary 
second order differential equation:
\begin{eqnarray}
\left (\partial_t-\lambda(t)\right ) \left [i \partial_t -L_0(t) \right ] \avX&=&
-i \Omega^2 \epsilon g^2(t) L_1^2 \avX,\nonumber\\ 
 i(\partial_t \avX)_0&=&L(0) \avX_0, \quad \avX_0=X_0;
\label{e9201} 
\end{eqnarray}
with $\lambda(t)\equiv -\epsilon+g'(t)/g(t)$.
Let's remark 
 that this equation is exact, 
unless the approximations we have seen  in the previous section.

\begin{figure}[p]
\centering\hspace{0.8cm}
\epsfig{file=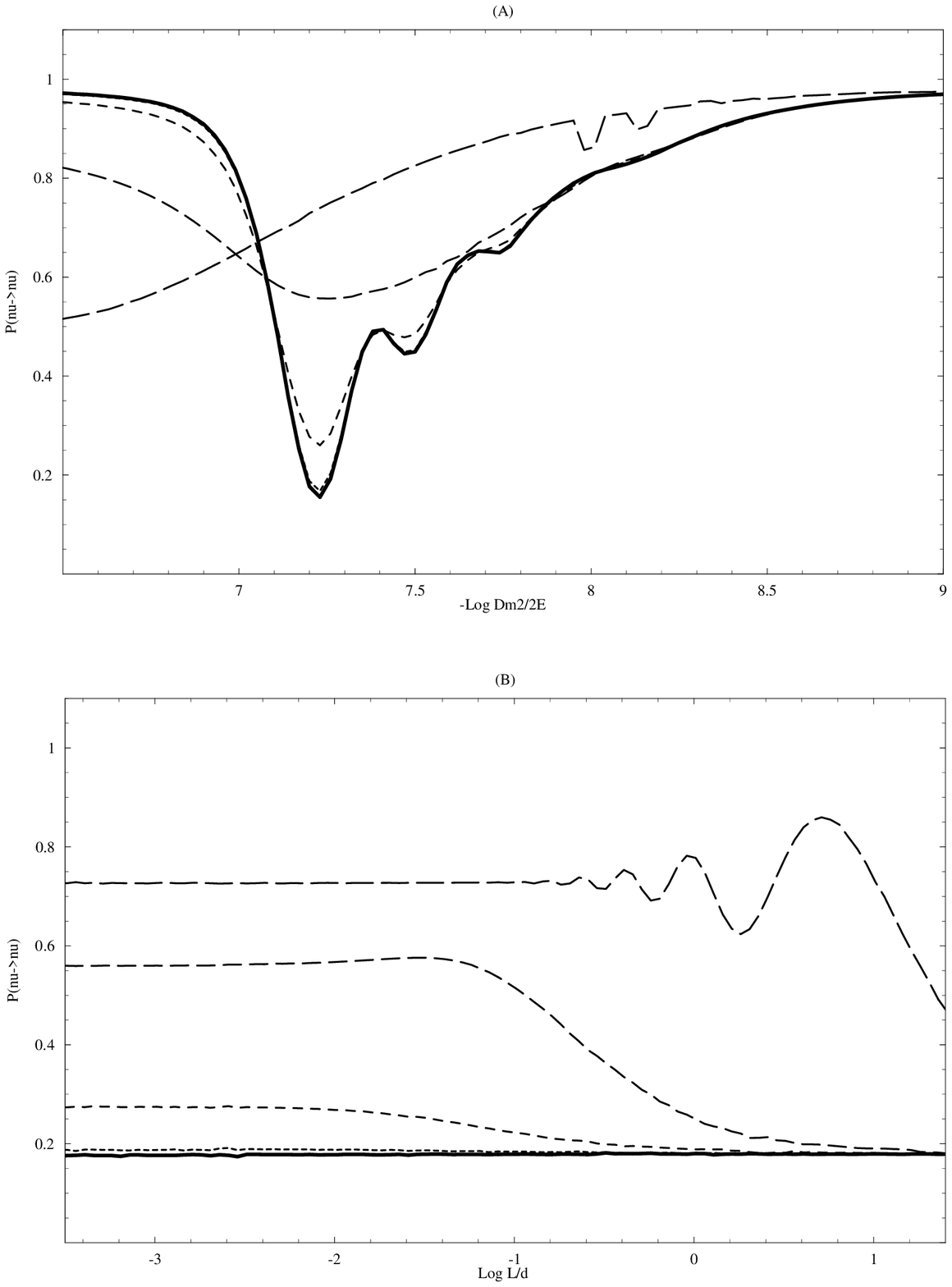,height=16cm}
\caption{ 
Neutrino oscillations in presence of random solar matter.
 (A) Averaged survival probability as a function of 
$\Delta m^2/2E$ ( $\delta$-correlation case, Eq.(\protect\ref{e6103} )).
Continuos curve: absence of noise. 
Dashed lines;
$\Omega^2=100,1,0.1,0.01\ (\times$ 1000 Km, from longer  to shorter dashing).
(B) The neutrino survival probability 
is plotted as a function  of $L/d$ with  $L\equiv 1/\epsilon$,
 ''d'' is the distance over which the chaotic field is acting 
($\approx 0.6\ R_\odot$). 
Eq.(\protect\ref{e9201}) is used for the same 
values of $\Omega^2$ as before (    
 $\Delta m^2/2 E=6.3\ 10^{-8}$ eV$^2/$MeV now).  
For both plots,
the neutrino, created at $r\approx 0.4\ R_\odot$ as a pure electron neutrino, 
propagates radially outwards. The flavour mixing angle is 
$\sin^2 2\theta=0.05$.}
\label{f1}
\end{figure}

As illustration and for the sake of comparation, 
we have solved Eqs.(\ref{e9201}) 
and (\ref{e6103}) for the case of  
two-flavour ultrarelativistic 
 neutrino oscillations  in presence of solar matter with 
a random perturbation. The function
$g(t)$ is in this case proportional to the solar electron density profile and
can be approximated rather well by an exponential function
 (see  notation and details in \cite{lor1} for example).
The results of the numerical calculations are shown in Figs.(\ref{f1}) (A)
(Eq.(\ref{e6103})) 
and (B) (Eq.(\ref{e9201})).
In plot (A) we observe the typical structure of the solar MSW resonance 
as a function of the neutrino energy and how this resonance is modified by 
the presence of the $\delta$ correlated chaotic field.
The broad range of validity of Eq.(\ref{e6103}), considered as an 
approximation to the more realistic Eq.(\ref{e9201}) is apparent in
plot (B). We observe there how the  survival 
probability depends on $\epsilon$. The neutrino energy is fixed in this 
case to be at the center of the resonance region and corresponds  
approximately to a neutrino oscillation length of $3.1\ 10^3$ Km, 
$4\ 10^{-3}$ R$_\odot$  or $8\ 10^{-3}$ times the distance over the 
which the noise is acting. For any fixed value of the quantity $\Omega^2$, 
the limit of small $\epsilon$ (large L/d in the figure) coincide 
with the noise-less  case. After a relatively narrow transition
 region, for larger, but in fact not so large, 
$\epsilon$ ( $L\approx d$) the probability 
tends to a constant; the one which is obtained from Eq.(\ref{e6103}).
The conclusion is that, at least for the example in consideration, the 
$\delta$-correlated noise is a good model even for ''not so small'' 
 correlation lengths.  
\footnote{
These results  should be compared with the ''cell''   calculations 
presented in \protect\cite{bur2}. In such approach the averaged probability is 
computed by direct ''Monte-Carlo'' methods. In order to 
simulate a finite correlation length  
 a step function correlation is used:
$f(t,t')$ is negligible whenever $\mid t-t'\mid$ is greater than some 
characteristic scale, otherwise is a constant.}

\section{Higher Order Moments and the Quantum Regression Theorem.} 
Second order  moments, expressions of the type
$\langle X_i X_j\rangle$, or in general, moments of any order, 
can be computed also using equations similar to  Eq.(\ref{e6102}). 
The straightforward procedure is to  
 define products $X_{ij..k}=X_i X_j.. X_k$ and write differential equations 
for them using the constitutive equations for each of the $X_i$. 
 The resulting equations are of the same type as  Eq.(\ref{e6101}).  
The similarity is   obvious when 
one adopts a tensorial notation and defines products of the form
$X^{(n)}=X\otimes...\otimes X^\dagger. $ The derivation of equations is specially simple within this notation.

In the computation of  averages of  
quantities of physical interest, i.e. expected signal rates,  
appear  also correlators of quantities at different
times. In the most simple case expressions 
of the type $\langle X_i(t+\tau) X_j(t)\rangle$.
We will show shortly that the knowledge of equal-time moments  and the application
of the quantum correlation theorem is sufficient for the computation of this
kind of correlations.

The  expected signal rate in a given experiment averaged
 over a finite range of energy 
and time can be described in general by a simplified
 expression of the form:
\begin{eqnarray}
N&=&\int_{\Delta E,\Delta t} dE dt\ \phi(E)\  P(E; t)
\label{e6109}
\end{eqnarray}
The function $\phi(E)$ is the experimental effective flux, where the original 
neutrino flux, the experimental detection and geometrical efficiencies and 
cross sections are included. 
A  hidden dependence, potentially relevant, on $E_0,t_0$, the central points of the integration intervals, is understood.

For a fluctuating survival probability $P(E;t)$,  $N$  is itself a random 
,non Gaussian, variable.  The averaged signal is simply the integral of the 
averaged probability. The second moment 
of the N distribution or 
more complicated energy or time correlations, i.e. averages of the type 
$$\left\langle N(E_0', t_0') N(E_0,t_0)\right \rangle $$ 
are all of the same form, schematically:
\begin{eqnarray}
\langle N^2\rangle &=&\int_{\Delta E,\Delta t} dE_{1,2}  dt_{1,2}  \phi(E_1)\phi(E_2) 
\langle P(E_1,t_1)P(E_2,t_2)\rangle
\label{e6110}
\end{eqnarray}

The correlation function appearing in the integrand of Eq.(\ref{e6110}) 
 can be computed 
in two steps. First 
we define the generalized density matrix $X^{(2)}$ as the tensorial product of 
usual density matrices at two different times and energies:
\begin{eqnarray}
X^{(2)}(E_1,E_2; t_1,t_2)&\equiv& X(E_1,t_1)\otimes X(E_2,t_2).
\label{e6111}
\end{eqnarray}
The  average of the 
element $\langle X^{(2)}_{1111}\rangle=\langle X_{11}X_{11}\rangle$ 
is evidently the probability correlation function we are looking 
for.

The differential equation for the equal time $X^{(2)}(E_1,E_2; t,t)$ is
obtained
 from the individual evolution equations for the matrices 
$X_{1,2}\equiv X(E_{1,2}) $ ( 
indices (1,2) label respectively expressions where $E_1,E_2$ appear)
\begin{eqnarray}
\partial_t X^{(2)}&\equiv& H_1 X^{(2)}+X^{(2)} H_2\equiv\rho L X^{(2)} 
\label{e6112}
\end{eqnarray}
Eq.(\ref{e6112}) is a random linear differential equation, linear in the 
stochastic variable $\rho(t)$. 
Applying the formalism developed in the previous 
section, we can write immediately the equation for the
ensemble average  $\langle X^{(2)}\rangle$ (Eq.(\ref{e6102}-\ref{e6103})).
Once we know the equal time correlator $ \langle P(t)P(t)\rangle$, we
obtain the expression for any other pair $(t,t')$ using the 
quantum regression theorem (\cite{gard1})
which reads as follows.
For any vector Markov process $Y$, if the ensemble average of $Y$ fulfills
a Schroedinger-like equation of the type
\begin{eqnarray}
\partial_t \langle Y(t)\rangle=G(t) \langle Y(t)\rangle,
\end{eqnarray}
with $G(t)$ an arbitrary matrix, then 
the second order correlations will obey the 
following equation  
\begin{eqnarray}
\partial_\tau \langle Y_i(t+\tau) Y_l(t)\rangle=\sum_j G_{ij}(\tau) \langle Y_j(t+\tau) Y_l(t)\rangle.
\end{eqnarray}

Note that 
the quantum regression theorem is in principle not applicable 
to systems described by the integral equation Eq.(\ref{e6102}). Nevertheless
 it is 
applicable to  problems where the integral equation can be reduced 
to a ordinary differential equation as Eq.(\ref{e9201}). Such second 
order equation can be  easily expressed as a first order one  
 by defining  the auxiliary pair process $Y=(X,\ \partial_t X)$.  

For the simplest case where $\rho$ is a $\delta$ correlated process
the equation for non-equal time correlators is explicitly given by:
\begin{eqnarray}
\partial_\tau \langle X^{(2)}(t+\tau,t)\rangle&=& -i \Omega^2 L^2(\tau)
\langle X^{(2)}(t+\tau,t)\rangle. 
\label{e6114}
\end{eqnarray}
For exponentially correlated problems second order  equations similar to 
Eq.(\ref{e9201}) can be immediately obtained.

\section{Conclusions and final Remarks.}
In conclusion, in this work 
we have  developed the general 
 formalism for the study of neutrino propagation in 
stochastic media. This formalism has allowed us 
 to derive systematically  evolution equations for averaged 
quantities as survival probabilities and higher order moments of them. 
New integro-differential equations
 valid for finite 
correlation processes have been  obtained for the first time. 
For exponentially correlated processes a second order ordinary equation is 
obtained as a consequence.
The Redfield equation valid
 for Gaussian delta-correlated
noise is rederived in  a simple way as a particular case. 
In the context for neutrino oscillations,
it has been shown that the presence of matter noise induces the appearance of 
a effective complex part in the matter density. 
The presence of magnetic 
field noise affects in a qualitatively different way. 
The wide range of validity of the $\delta$-correlated model 
has been checked in 
a realistic   example corresponding neutrino oscillations in random 
 solar matter. 
Finally,  the formalism together
with the quantum correlation theorem  has been  used to obtain 
equations for higher distribution moments and  non-equal time   
correlation functions.

The formalism can be generalized in an obvious way to 
perform
  the ensemble average of equations of  slightly more general type
 than Eq.(\ref{e6101}), equations of the form:
\begin{eqnarray}
i\partial_t X &=& \sum_i\rho_i(t) L_i(t) X\; ; \quad X(0)=X_0.
\label{e4101} 
\end{eqnarray}
These equations  appear for example in chaotic neutrino magnetic precession.
The result for the case where the $\rho_i$ are $\delta$-correlated in time but  mutually uncorrelated is simply: 
\begin{eqnarray}
i\partial_t \avXt&=&-i \sum_i\Omega_i^2 L_i^2(t)\avXt.
\end{eqnarray}

The generalization to continuous Schroedinger equations in presence of 
random potentials or random external forces is also obvious if we left aside 
the mathematical differences coming from the appearance of infinite
dimensional Hilbert spaces.
The formalism is of general application to any 
quantum system governed by similar Schroedinger equations: equations where
a stochastic function appears multiplicatively in some term of the 
Hamiltonian.

\vspace{1cm}

{\bf Acknowledgments}. 

 This work has been supported by DGICYT under Grant 
 PB95-1077 and by  a DGICYT-MEC contract  at Univ. de Valencia. 
Early versions of this work were developed at the Institute Fur 
 Theoretische Physik, Universitat Bern, supported by a grant 
from the Wolferman-Nageli Foundation.

\newpage



\end{document}